\documentclass[aps,prx,twocolumn,superscriptaddress,longbibliography]{revtex4-1}
\usepackage{graphicx}
\usepackage{dcolumn}
\usepackage{bm}
\usepackage{color}
\usepackage[usenames,dvipsnames,svgnames,table]{xcolor}
\usepackage[utf8]{inputenc}
\usepackage[english]{babel}

\newcommand{\LSCO}{La$ _{2-x} $Sr$ _{x} $CuO$ _{4}$}
\newcommand{\RH}{$R_{\rm H}$}

\newcommand{\Tcdw}{$T_{\rm CDW}$}
\newcommand{\Tsdw}{$T_{\rm SDW}$}
\newcommand{\Tmax}{$T_{\rm max}$}
\newcommand{\Tc}{$T_{\rm c}$}
\newcommand{\pcdw}{$p_{\rm CDW}$}
\newcommand{\pfsr}{$p_{\rm FSR}$}

\begin{document}


\title{
%
Critical doping for the onset of Fermi-surface reconstruction by charge-density-wave order
in the cuprate superconductor La$ _{2-x} $Sr$_{x} $CuO$ _{4}$
}



\author{S.~Badoux}
\email[]{sven.badoux@usherbrooke.ca}
\affiliation{D\'epartement de physique \& RQMP, Universit\'e de Sherbrooke, Sherbrooke, Qu\'ebec J1K 2R1, Canada}

\author{S.A.A.~Afshar} 
\affiliation{D\'epartement de physique \& RQMP, Universit\'e de Sherbrooke, Sherbrooke, Qu\'ebec J1K 2R1, Canada}

\author{B.~Michon}
\affiliation{D\'epartement de physique \& RQMP, Universit\'e de Sherbrooke, Sherbrooke, Qu\'ebec J1K 2R1, Canada}

\author{A.~Ouellet} 
\affiliation{D\'epartement de physique \& RQMP, Universit\'e de Sherbrooke, Sherbrooke, Qu\'ebec J1K 2R1, Canada}

\author{S.~Fortier} 
\affiliation{D\'epartement de physique \& RQMP, Universit\'e de Sherbrooke, Sherbrooke, Qu\'ebec J1K 2R1, Canada}

\author{D.~LeBoeuf}
\affiliation{Laboratoire National des Champs Magn\'etiques Intenses, UPR 3228, (CNRS-INSA-UJF-UPS), Grenoble 38042, France}

\author{T.P.~Croft}
\affiliation{H. H. Wills Physics Laboratory, University of Bristol, Bristol, BS8 1TL, United Kingdom}

\author{C.~Lester}
\affiliation{H. H. Wills Physics Laboratory, University of Bristol, Bristol, BS8 1TL, United Kingdom}

\author{S.M.~Hayden}
\affiliation{H. H. Wills Physics Laboratory, University of Bristol, Bristol, BS8 1TL, United Kingdom}

\author{H.~Takagi} 
\affiliation{Department of Advanced Materials, University of Tokyo, Kashiwa 277-8561, Japan}

\author{K.~Yamada} 
\affiliation{Institute of Materials Structure Science, High Energy Accelerator Research Organization
\& The Graduate University for Advanced Studies, Oho, Tsukuba 305-0801, Japan}

\author{D.~Graf} 
\affiliation{National High Magnetic Field Laboratory, Tallahassee, FL 32310, USA}

\author{N.~Doiron-Leyraud}
\affiliation{D\'epartement de physique \& RQMP, Universit\'e de Sherbrooke, Sherbrooke, Qu\'ebec J1K 2R1, Canada}

\author{Louis~Taillefer}
\email[]{louis.taillefer@usherbrooke.ca}
\affiliation{D\'epartement de physique \& RQMP, Universit\'e de Sherbrooke, Sherbrooke, Qu\'ebec J1K 2R1, Canada}
\affiliation{Canadian Institute for Advanced Research, Toronto, Ontario M5G 1Z8, Canada}
\date{\today}

\begin{abstract}

The Seebeck coefficient $S$ of the cuprate superconductor \LSCO~(LSCO) was measured in magnetic fields large enough to 
access the normal state at low temperatures,
for a range of Sr concentrations from $x = 0.07$ to $x = 0.15$. 
For $x = 0.11$, 0.12, 0.125 and 0.13, $S/T$ decreases upon cooling to become negative at low temperatures. 
The same behavior is observed in the Hall coefficient \RH $(T)$. 
In analogy with other hole-doped cuprates at similar hole concentrations $p$, the negative $S$ and \RH~show that the Fermi surface of LSCO undergoes a reconstruction caused by the onset of charge-density-wave modulations. 
Such modulations have indeed been detected in LSCO by X-ray diffraction in precisely the same doping range. 
Our data show that in LSCO this Fermi-surface reconstruction is confined to $0.085 < p < 0.15$. 
We argue that in the field-induced normal state of LSCO, 
charge-density-wave order ends at a critical doping $p_{\rm CDW} = 0.15 \pm 0.005$, well below the pseudogap critical doping $p^\star \simeq 0.19$.

\end{abstract}


\maketitle




Since the discovery of quantum oscillations \cite{NDL2007} and a negative Hall coefficient \RH~\cite{LeBoeuf2007} in the cuprate superconductor 
YBa$_2$Cu$_3$O$_y$ (YBCO), it has become clear that the Fermi surface of underdoped YBCO undergoes a reconstruction at low temperature that produces a small electron pocket \cite{Taillefer2009}, in a doping range from $p = 0.08$ to $p \simeq 0.15$ \cite{LeBoeuf2011}. 
This Fermi-surface reconstruction (FSR) was also detected as a sign change in the Seebeck coefficient $S(T)$, going from positive at high temperature to negative at low temperature \cite{Chang2010}. 
A strikingly similar change of sign in $S(T)$ observed in the cuprate La$_{1.8-x}$Eu$_{0.2}$Sr$_x$CuO$_4$ (Eu-LSCO) \cite{Laliberte2011} suggested that the stripe order known to exist in Eu-LSCO \cite{Fink2011} -- a combination of charge-density-wave (CDW) and spin-density-wave (SDW) modulations -- is responsible for the FSR in both materials. 
The observation of CDW modulations in YBCO by NMR \cite{Wu2011} and X-ray diffraction (XRD)  \cite{Ghiringhelli2012, Chang2012} 
confirmed this conjecture, and demonstrated that it is the CDW (and not the SDW) modulations that cause the FSR.

In YBCO, the drop in \RH $(T)$ and $S/T$ begins at a temperature \Tmax~that peaks at $p = 0.12$ (Fig.~1a). 
The drop is attributed to the CDW modulations detected by XRD \cite{Hucker2014,Blanco-Canosa2014} and NMR \cite{Wu2015} below a temperature \Tcdw~in the same doping range as the FSR \cite{LeBoeuf2011}, with \Tcdw~also peaking at $p = 0.12$ (Fig.~1a).

In HgBa$_2$CuO$_{4 + \delta}$ (Hg1201), high-field measurements of Hall and Seebeck coefficients revealed a similar FSR \cite{NDL2013}, confirmed by the observation of quantum oscillations \cite{Barisic2013} and again attributed to XRD-detected CDW modulations \cite{Tabis2014}. 
All this suggests that CDW modulations and the associated FSR are generic properties of hole-doped cuprates in the vicinity of $p = 0.12$. 
A major outstanding question is : Up to what critical doping \pcdw~do CDW modulations extend in the phase diagram (Fig.~1),
in particular in the field-induced normal state 
at $T=0$ ?
In this context, the material LSCO offers a powerful platform, since good crystals can be grown with $p$ up to 0.3 and beyond.
CDW modulations have been observed in LSCO with XRD, at $p \simeq 0.12$ \cite{Croft2014, Thampy2014, Christensen2014},
but there is little information about the associated FSR.


In this Article, we report high-field measurements of the Seebeck coefficient in LSCO single crystals at several dopings, which show that $S$ becomes 
negative in the normal state at low temperature in precisely the doping range where CDW modulations are detected by XRD. 
\RH~is also found to be negative in that range.
The FSR in LSCO is therefore very similar to the FSR in YBCO and Hg1201. 
Our data show that the FSR does not extend above $p=0.15$, strong evidence that CDW order in LSCO ends at a critical doping \pcdw~$= 0.15$.
This implies that in the normal state of LSCO the phase of CDW order ends well before the pseudogap phase,
which ends at the critical doping $p^\star \simeq 0.19$ (ref.~\onlinecite{Cooper2009}).

 %


\begin{figure}[h!]
\includegraphics[scale=0.66]{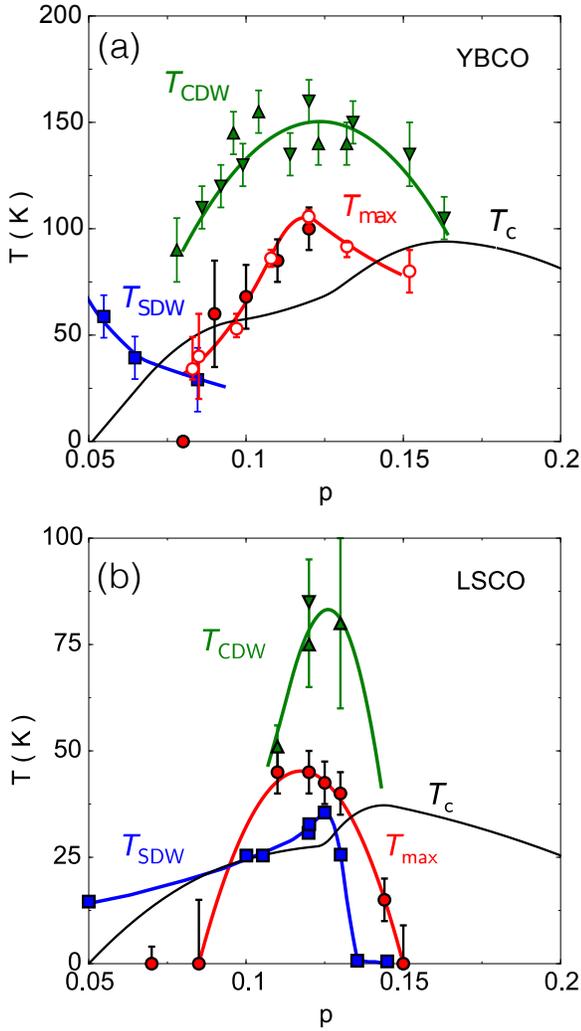}
\caption{ 
Temperature-doping phase diagram of the cuprate superconductors YBCO (a) and LSCO (b).
The superconducting transition temperature \Tc~is drawn as a black line.
Charge density-wave (CDW) modulations are detected by X-ray diffraction below
\Tcdw~(green triangles) in YBCO (up triangles \citep{Hucker2014}, down triangles \citep{Blanco-Canosa2014}) and LSCO (up triangles \citep{Croft2014}, 
down triangle \citep{Christensen2014}). 
Spin-density-wave (SDW) modulations are detected by neutron diffraction below \Tsdw~(blue squares) in YBCO \citep{Haug2010} and LSCO \cite{Chang2008,Kofu2009,Wakimoto1999,Lake2002,Kimura1999}. 
When plotted as $S/T$ vs $T$, the normal-state Seebeck coefficient peaks at a temperature \Tmax~(full red circles) 
before it drops at low temperature due to Fermi-surface reconstruction (YBCO, ref.~\onlinecite{Laliberte2011}; LSCO, this work, Figs.~3 and 4).
A similar \Tmax~can also be defined for the Hall coefficient (open red circles), below which \RH$(T)$ drops at low temperature (YBCO, ref.~\onlinecite{LeBoeuf2011}).
}
%
\end{figure}



\begin{figure}
\includegraphics[width=8.9cm]{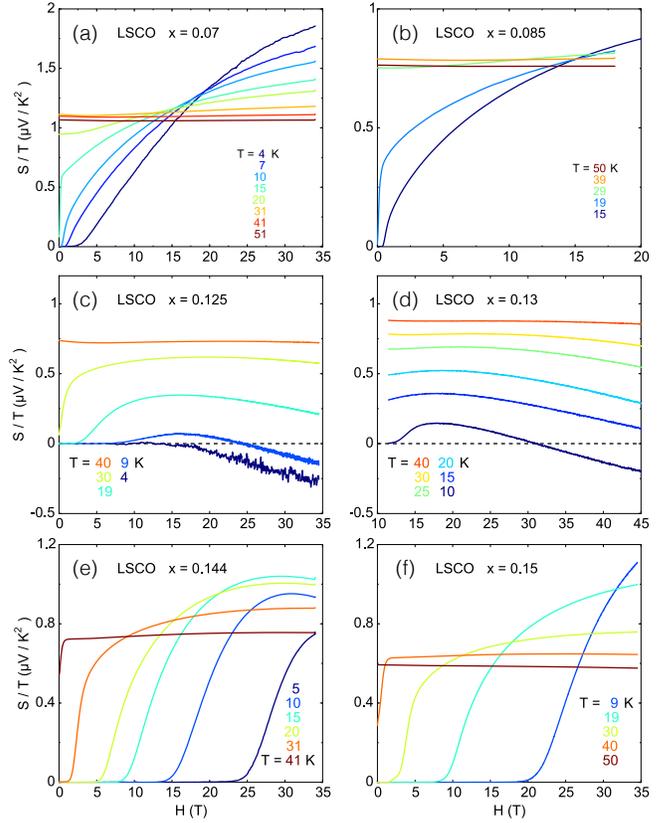}
\caption{ 
Isotherms of the Seebeck coefficient in LSCO, plotted as $S/T$ vs magnetic field $H$,
at various temperatures, as indicated, for six samples, with 
$x = 0.07$ (a), 
$x = 0.085$ (b),
$x = 0.125$ (c), 
$x = 0.13$ (d),
$x = 0.144$ (e),
and $x = 0.15$ (f).
For $x = 0.125$ and 0.13, $S/T$ at high $H$ decreases at low temperature, to reach negative values.
For $x = 0.144$, $S/T$ also decreases at low temperature, below 15~K.
This decrease is the signature of FSR.
In contrast, for $x = 0.07$ and 0.15, $S/T$ at the highest measured field keeps increasing with decreasing temperature down 
to the lowest temperature.
This shows that there is no FSR at those dopings, at least down to 4~K and 9~K, respectively.
The same is true at $x = 0.085$, at least down to 15~K.
}
%
\end{figure}



\begin{figure}[h!]
\includegraphics[width=8.8cm]{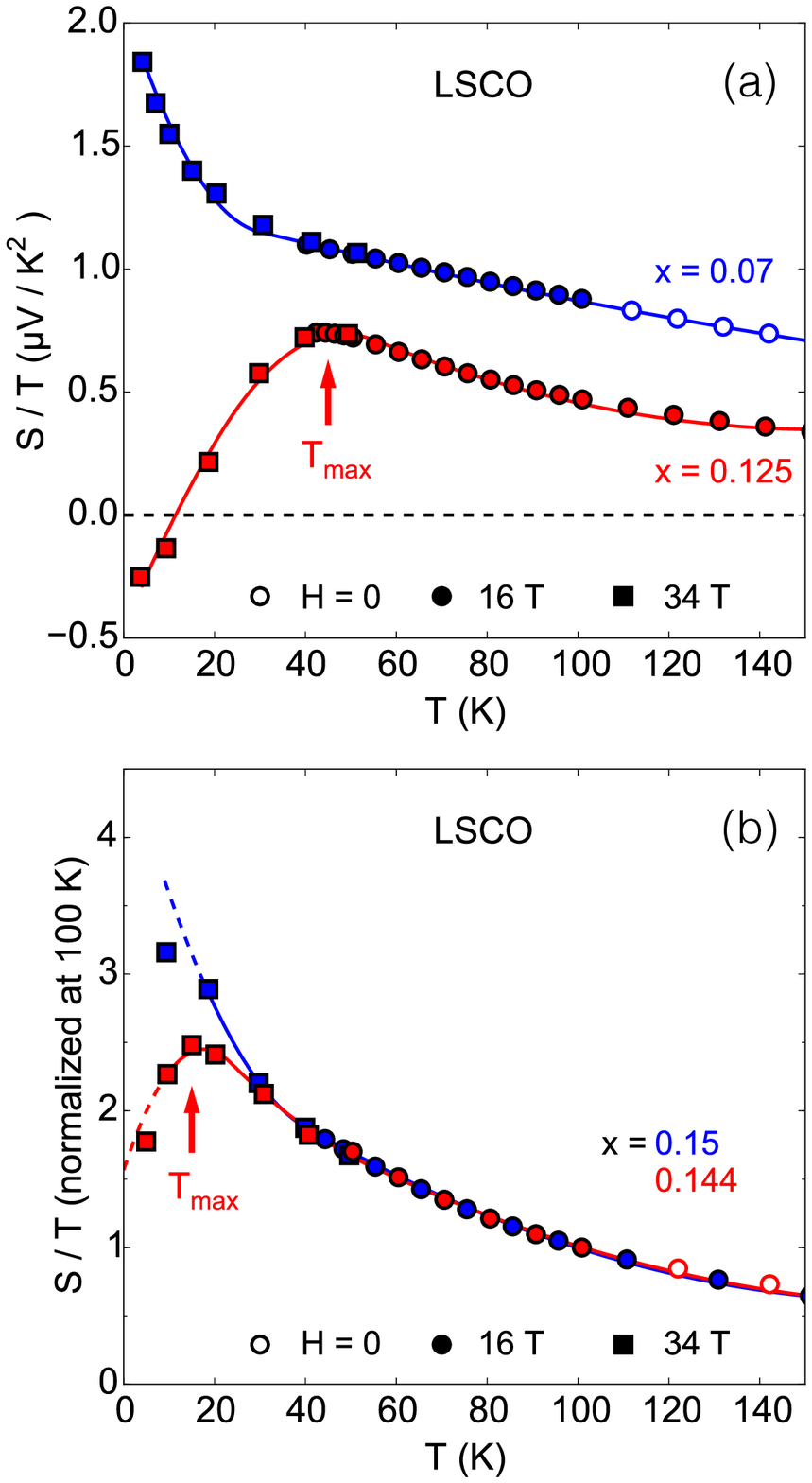}
\caption{ 
Seebeck coefficient of LSCO, plotted as $S/T$ vs temperature $T$,
measured in a magnetic field $H = 0$ (open circles), 16~T (full circles) and 34~T (squares), 
for four samples, with 
$x = 0.07$ and 0.125 (a),
and $x = 0.144$ and 0.15 (b).
The data in panel b are normalized to their value at $T = 100$~K.
All data points represent the normal state, for which the solid lines are a guide to the eye,
except the lowest point for each of $x = 0.144$ and $x = 0.15$ (panel b).
For these two points, the isotherms are still going up the superconducting transition (Fig.~2).
The dashed lines are an extension of the normal-state behavior based on extrapolating those isotherms beyond 34~T.
\Tmax~marks the temperature below which $S/T$ decreases at low temperature (arrow),
in some cases to reach negative values, as seen here for $x = 0.125$.
This decrease is the signature of Fermi-surface reconstruction (FSR).
Note how the data for $x = 0.144$ and $x = 0.15$ split
below $T \simeq 30$~K,
with the former dropping at low $T$ due to FSR and the latter showing no decrease, and hence no FSR (at least down to 9~K).
}
\end{figure}



{\it Methods.--}
Single crystals of LSCO were grown by the flux-zone technique with Sr concentrations $x = 0.085$, 0.11, 0.12 and 0.13 at the University of Bristol, 
$x = 0.07$ and 0.125 at the University of Tokyo, $x=0.144$ and 0.15 at Tohoku University. 
Samples were cut in the shape of rectangular platelets, with typical dimensions 0.5~mm $\times$ 1.0~mm $\times~0.1$~mm. 
The hole concentration (doping) $p$ is taken to be $p = x$. 
The (zero-resistance) superconducting transition temperature of the 8 samples is \Tc~$= 12.7$, 20.2, 26.2, 27.5, 28.0, 32.3, 37.2, and 36.5~K for $p = 0.07$, 
0.085, 0.11, 0.12, 0.125, 0.13, 0.144, and 0.15, respectively.
The Seebeck coefficient was measured, as described elsewhere \cite{Laliberte2011}, at Sherbrooke (all samples) up to $H = 20$~T, 
at the National High Magnetic Field Laboratory (NHMFL) in Tallahassee up to $H = 34$~T ($x = 0.125$ and 0.15) 
and up to $H = 45$~T ($x = 0.13$), and at the Laboratoire National des Champs Magn\'{e}tiques Intenses (LNCMI) in Grenoble up to $H = 34$~T ($x = 0.07$ and 0.144).
The Hall coefficient of samples with $x = 0.11$, 0.12, 0.125 and 0.13 was measured, as described elsewhere \cite{LeBoeuf2011}, at Sherbrooke in $H = 16$~T. 
All crystals have an orthorhombic crystal structure and they are twinned.
The thermal gradient or electrical current was applied in the basal plane,
while the magnetic field was applied along the $c$ axis.


\begin{figure}
\includegraphics[width=8.8cm]{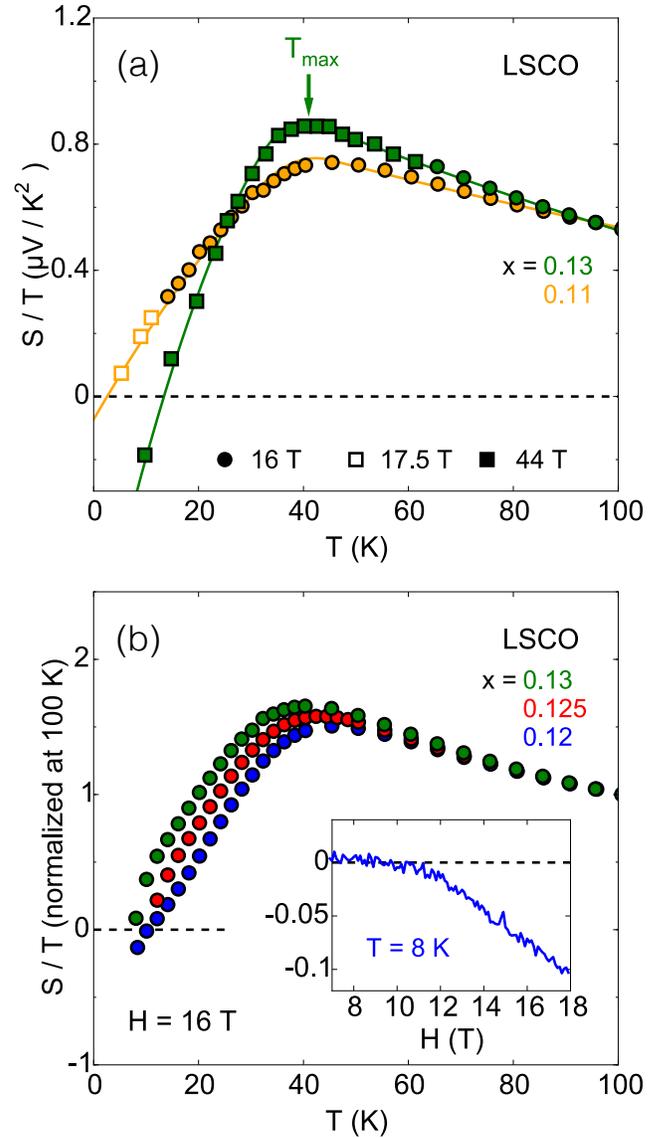}
\caption{ 
Same as in Fig.~3, for samples with $x = 0.11$ (yellow), $x = 0.12$ (blue), $x = 0.125$ (red) and $x = 0.13$ (green), 
measured at $H = 16$~T (full circles), 17.5~T (open squares) and 44~T (full squares).
The data in panel b are normalized to their value at T = 100 K.
FSR is clearly observed in all four samples, as a drop in $S/T$ at low temperature.
{\it Inset of panel b}:
Isotherm at $T = 8$~K for $x = 0.12$, showing that $S/T$ becomes increasingly negative with increasing field,
demonstrating that the negative $S$ is a property of the normal state.
}
\end{figure}



\begin{figure*}
\includegraphics[scale=0.549]{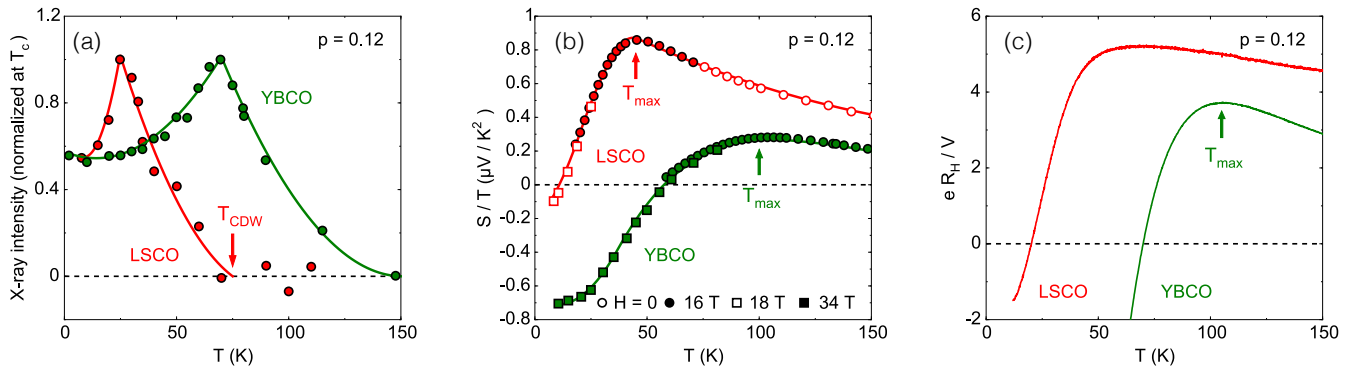}
\caption{
Comparison of LSCO (red) and YBCO (green) at $p = 0.12$.
a) 
Temperature dependence of the X-ray intensity associated with the CDW modulations, normalized at \Tc, 
detected in LSCO \cite{Croft2014} and YBCO \cite{Chang2012}.
Lines are a guide to the eye.
The cusp is at \Tc.
b) Normal-state Seebeck coefficient of LSCO (this work) and YBCO \cite{Laliberte2011}, measured in a magnetic field as indicated,
plotted as $S/T$ vs $T$. 
\Tmax~is the temperature below which $S/T$ drops to reach negative values at low temperature (arrow), 
the signature of Fermi-surface reconstruction (FSR).
This \Tmax~is plotted as full circles in Fig.~1.
Lines are a guide to the eye.
c) 
Hall coefficient of LSCO at $H = 16$~T and YBCO at $H = 15$~T \cite{LeBoeuf2007},
plotted as $e$\RH$/V$, where $e$ is the electron charge and $V$ the volume per planar Cu atom. 
\Tmax~is the temperature below which $R_{\rm H}(T)$ drops to reach negative values at low temperature (arrow), 
another signature of FSR.
\Tmax~is plotted as open circles in Fig.~1a \cite{LeBoeuf2011}.
}
\end{figure*}



{\it Seebeck coefficient.--}
In Fig.~2, the Seebeck data for 6 samples are plotted as $S/T$ vs $H$ for several temperatures. 
We see that for $x = 0.125$ (Fig.~2c) and $x = 0.13$ (Fig.~2d), $S$ becomes negative at high field and low temperature.
%
This shows that a negative $S$ is a property of the normal state of LSCO at these dopings, 
as in YBCO, Eu-LSCO and Hg1201.
At $x = 0.144$, we see that at high field $S/T$ decreases when the temperature drops below $T = 15$~K (Fig.~2e).
In contrast, no such decrease is observed at $x = 0.15$, down to the lowest temperature (Fig.~2f).
At $x = 0.07$, $S/T$ increases steadily with decreasing $T$ at high field, down to the lowest temperature (Fig.~2a).
This is also true at $x = 0.085$ (Fig.~2b).
Although here our data only goes to 20~T, the crossing of the lowest isotherms shows that 
$S/T$ keeps increasing down to $T = 15$~K, at least.

In Figs.~3 and 4, we plot $S/T$ vs $T$, at high field.
In Fig.~3a, we see that the drop in $S/T$ at $x = 0.125$ to negative values starts below a temperature \Tmax~$\simeq 40$~K.
This is also the case at $x = 0.11$ and 0.13 (Fig.~4a). 
In Fig.~4b, we compare data on 3 samples taken in identical conditions, at $H = 16$~T.
(Although the LSCO sample with $x = 0.12$ was only measured
up to 18~T, $S/T$ at $T = 8$~K is increasingly negative with increasing $H$ (inset of Fig.~4b),
confirming that a negative $S$ is a property of the normal state also at that doping.)
The location of the peak in $S/T$ vs $T$ is seen to decrease from \Tmax~$=45$~K at $x = 0.12$, 
to \Tmax~$=42.5$~K at $x = 0.125$, to \Tmax~$=40$~K at $x = 0.13$.
Those \Tmax~values are plotted on the phase diagram of LSCO in Fig.~1b.
Raising the doping further, we observe that \Tmax~continues its steady descent.
Indeed, at $p = 0.144$, $S/T$ now peaks at \Tmax~$\simeq 15$~K (Fig.~3b).
Extrapolating this trend yields \Tmax~$\to 0$ at $p \to 0.15$ (Fig.~1b).
Our data at $x = 0.15$ confirm this, with $S/T$ showing no decrease down to
at least 9~K (Figs.~2f and 3b). 
This shows that FSR in LSCO ends at a critical doping \pfsr~$= 0.15 \pm 0.005$.

At $x = 0.07$, the normal-state $S/T$ increases monotonically with decreasing $T$, down to our lowest temperature (Fig.~3a).
There is clearly no FSR at that doping.
At $x = 0.085$, although we only measured up to 18 or 20~T, we observe that $S/T$ at $H = 18$~T
increases as $T \to 0$, at least down to 15~K (Fig.~2b).
So here \Tmax~$< 15$~K.
In Fig.~1b, we plot \Tmax~vs $p$ for our 8 samples, with their uncertainty, and thereby delineate the region where FSR occurs in the $T-p$ phase diagram
of LSCO. 
We see that the FSR region peaks at $p \simeq 0.12$ and is confined between $p \simeq 0.085$
and $p =$ \pfsr~$= 0.15 \pm 0.005$.\\


{\it Hall coefficient.--}
In Fig.~5c, the Hall coefficient of our LSCO crystal with $x = 0.12$, measured at $H = 16$~T, is plotted as \RH~vs $T$. 
We see that \RH $(T)$ drops below $T~\simeq~50$~K and becomes negative below $T \simeq 20$~K. 
Data for our crystals with $x = 0.11$, 0.125 and 0.13 are very similar, also negative at low $T$,
all in excellent agreement with prior low-field data on single crystals of LSCO with $x = 0.12$ \cite{Suzuki2002}.
(The absence of a negative \RH~in previous high-field data on thin films of LSCO \cite{Balakirev2009}
may be due to the higher disorder of such samples.)
A similar drop in \RH$(T)$~has been seen in Eu-LSCO \cite{Cyr-Choiniere2009} and in
La$_{1.4-x}$Nd$_{0.6}$Sr$_x$CuO$_4$ (Nd-LSCO) \cite{Noda1999}, when $p \simeq 0.12$;
in both materials, it is closely linked to the onset of CDW order.\\


{\it Discussion.--}
Taken together, the negative Hall and Seebeck coefficients in the normal state of LSCO are conclusive evidence of FSR in this material, in the vicinity of $p = 0.12$.
This adds to the previous three cases, namely YBCO, Eu-LSCO and Hg1201. 
In all 4 cases, the FSR occurs in a region of the $T-p$ phase diagram where CDW modulations have been detected by XRD (Fig.~1). 
The link between CDW and FSR is robust.

It is instructive to compare LSCO and YBCO. 
The two phase diagrams are similar (Fig.~1). 
In both cases, \Tmax~and \Tcdw~peak at $p = 0.12$, and the region of FSR is confined to similar ranges -- from $p \simeq 0.085$ to $p = 0.15$ in LSCO and from $p = 0.08$ to $p \simeq 0.15$ in YBCO \cite{LeBoeuf2011}. 
In Fig.~5, we compare data for LSCO and YBCO directly, at $p = 0.12$. 
The CDW modulations detected by XRD emerge below a temperature twice as high in YBCO compared to LSCO (Fig.~5a) : 
\Tcdw~$\simeq 150$~K in YBCO vs \Tcdw~$\simeq 75$~K in LSCO. 
Correspondingly, the FSR is detected at a temperature twice as high in YBCO compared to LSCO, 
with \Tmax~$\simeq 100$~K in YBCO vs \Tmax~$\simeq 50$~K in LSCO (Fig. 5b). 
All this suggests that CDW ordering is a stronger tendency in YBCO than in LSCO. 
Intriguingly, the superconducting transition temperature \Tc~is roughly twice as high in YBCO as compared to LSCO (see cusp in Fig.~5a). 
This raises the interesting possibility that the same underlying mechanism, perhaps magnetic, fuels both superconductivity and CDW order \cite{Efetov2013}.

Given that FSR in LSCO ends at \pfsr~$= 0.15$, we infer that this is also where CDW order ends.
This is consistent with recent XRD measurements that detect no CDW modulations in LSCO at $x = 0.15$ \cite{Croft-Hayden-private}.
(The same consistency is observed at $x = 0.085$, where again no CDW modulations are detected by XRD \cite{Croft-Hayden-private}.)
We thus arrive at a key information : the CDW phase in LSCO ends at the critical doping \pcdw~$= 0.15$.

This is distinctly below the critical point where the pseudogap phase is believed to end in LSCO,
at $p^\star \simeq 0.19$, as determined from the normal-state resistivity measured in high magnetic fields \cite{Cooper2009}.
This clear separation reveals that the pseudogap phase is not caused by the CDW ordering.
Instead, it suggests that CDW order is a secondary instability of the pseudogap phase.
A very similar separation was recently observed in YBCO from high-field Hall effect measurements,
with \pcdw~$= 0.16 \pm 0.005$ and $p^\star \simeq 0.19$ \cite{Badoux2015}.
This strongly suggests that a separation of \pcdw~and $p^\star$ 
is a generic property of cuprates.

%
%
%
%


{\it Summary.--}
Our high-field measurements of the Seebeck coefficient in the cuprate superconductor LSCO reveal that its normal-state Fermi surface undergoes a reconstruction at low temperature, in the doping range $0.085 < p < 0.15$. 
In analogy with the cuprates YBCO, Eu-LSCO and Hg1201, we attribute this FSR to the CDW modulations detected by XRD in the very same doping range. 
Combined with XRD data on LSCO, our Seebeck data make a compelling case that CDW modulations disappear at $p = p_{\rm CDW} = 0.15$, 
so that the field-induced non-superconducting ground state of LSCO above $p = 0.15$
has no CDW order.
Because the pseudogap phase in the normal state of LSCO extends up to $p \simeq 0.19$,
we infer that the pseudogap is not tied to CDW ordering.
Instead, the CDW modulations appear to be a secondary instability of the pseudogap phase.

%
%


{\it Acknowledgements.--}
We thank J.~Chang, N.~E.~Hussey, M.-H.~Julien, P.~A.~Lee, B.~J.~Ramshaw, S.~Sachdev, J.~L.~Tallon, and G.~A.~Sawatzky for stimulating discussions.
A portion of this work was performed at the National High Magnetic Field Laboratory, which is supported by the National Science Foundation Cooperative Agreement No. DMR- 1157490, the State of Florida, and the U.S. Department of Energy. 
Another portion of this work was performed at the Laboratoire National des Champs Magn\'etiques Intenses of the CNRS, 
member of the European Magnetic Field Laboratory.
L.T. acknowledges support from the Canadian Institute for Advanced Research (CIFAR) and funding from the National Science and Engineering Research Council of Canada (NSERC), the Fonds de recherche du Qu\'ebec - Nature et Technologies (FRQNT), the Canada Foundation for Innovation (CFI) and a Canada Research Chair.


\bibliography{FSR_LSCO}

\end{document}